\documentclass[aps,prb,superscriptaddress,reprint]{revtex4-1} % preprint,reprint,

\usepackage[colorlinks=true,linkcolor=blue,citecolor=blue,urlcolor=blue]{hyperref}
\usepackage{setspace} %to set 1.5 spacing in text
\usepackage{graphicx}
\usepackage{amsmath}
\usepackage{color}
\usepackage{amsmath}
\usepackage{amssymb}
\usepackage{verbatim}
\usepackage{latexsym}
\usepackage{enumerate} % to change the numbering of 'enumerate'
\usepackage{bm} % for bold face mathematical symbols
\usepackage{ulem} %package to cross-out and highlight passages
\begin{document}

\title{
%Negative refraction of phonon surface Weyl arcs at interfaces of chiral $\alpha$-quartz crystal twins
Negative refraction of Weyl phonons at twin quartz interfaces
% Negative efraction of surface Weyl arc phonons at chiral $\alpha$-quartz crystal twins interfaces
%Negative refraction of chiral topological surface phonons at interfaces of $\alpha$-quartz crystal twins
%phonon surface Weyl arcs
%Negative refraction of chiral topological phonon isofrequency contours at surfaces of $\alpha$-quartz crystal twins
%{\color{red} (update SI once finishing the title!)}
}

\author{Juan D. F. Pottecher}
\thanks{These authors contributed equally to this work.}
\affiliation{St Catharine's College, University of Cambridge, Trumpington Street, Cambridge CB2 1RL, United Kingdom}

\author{Gunnar F. Lange}
\thanks{These authors contributed equally to this work.}
\affiliation{Theory of Condensed Matter Group, Cavendish Laboratory, University of Cambridge, J.\,J.\,Thomson Avenue, Cambridge CB3 0HE, United Kingdom}

\author{Cameron Robey}

\affiliation{St John's College, University of Cambridge, St John's Street, Cambridge CB2 1TP, United Kingdom}

\author{Bartomeu Monserrat}

\email{bm418@cam.ac.uk}

\affiliation{Theory of Condensed Matter Group, Cavendish Laboratory, University of Cambridge, J.\,J.\,Thomson Avenue, Cambridge CB3 0HE, United Kingdom}

\affiliation{Department of Materials Science and Metallurgy, University of Cambridge, 27 Charles Babbage Road, Cambridge CB3 0FS, United Kingdom}

\author{Bo Peng}

\email{bp432@cam.ac.uk}

\affiliation{Theory of Condensed Matter Group, Cavendish Laboratory, University of Cambridge, J.\,J.\,Thomson Avenue, Cambridge CB3 0HE, United Kingdom}

\date{\today}

\begin{abstract}
In nature, $\alpha$-quartz crystals frequently form contact twins - two adjacent crystals with the same chemical structure but different crystallographic orientation, sharing a common lattice plane. As $\alpha$-quartz crystallises in a chiral space group, such twinning can occur between enantiomorphs with the same handedness or with opposite handedness. Here, we use first-principle methods to investigate the effect of twinning and chirality on the bulk and surface phonon spectra, as well as on the topological properties of phonons in $\alpha$-quartz. We demonstrate that, even though the dispersion appears identical for all twins along all high-symmetry lines and at all high-symmetry points in the Brillouin zone, the dispersions can be distinct at generic momenta for some twin structures. Furthermore, when the twinning occurs between different enantiomorphs, the charges of all Weyl nodal points flip, which leads to mirror symmetric isofrequency contours of the surface arcs. We show that this allows negative refraction to occur at interfaces between certain twins of $\alpha$-quartz.
\end{abstract}

%\flushbottom

\maketitle

\section{Introduction}

% Negative refraction

Negative refraction is a counter-intuitive phenomenon in which incident and refracted waves emerge on the same side of the interface normal\,\cite{Veselago1968}, providing potential applications in superlens and sub-wavelength imaging\,\cite{Pendry2000}. %{\color{red} Negative refraction can be found in materials with antiparallel group and phase velocities\,\cite{Cheianov2007,Lezec2007,Lee2015a,Zhang2016f,Betancur-Ocampo2018,Lv2018}, with graphene p-n junctions providing an example in which the group and phase velocities are parallel or antiparallel for the electron-doped (n) and hole-doped (p) states respectively\,\cite{Cheianov2007,Lee2015a}.}
One prominent strategy to obtain negative refraction uses open isofrequency contours, which has been realised in hyperbolic metamaterials\,\cite{Lin2017,Zanotto2022,Hu2023,Sternbach2023}. Recent advances in topological materials offer a new platform to manipulate isofrequency contours arising from topological surface states\,\cite{Burkov2011,Fang2015,Sun2015,Huang2015b,Deng2016,Tamai2016,Fang2016a,Kuo2019,Wang2022a,Guo2023}. For example, the surface arcs of Weyl points can form distinct isofrequency contours for both positive and negative refraction\,\cite{He2018,Chen2020a},
%Geometry of gyroid acoustic crystals. a) Unit cells of gyroid surface with F = 0 and F > 0. b) Illustration of the inherent chirality of the gyroid surface. The red and black spiral lines indicate the left-handed helices and right-handed helices
and all-angle reflectionless negative refraction has been observed in Weyl metamaterials\,\cite{Liu2022}. %{\color{red} However, 
%although it has been proposed that in Weyl semimetals negative refraction can take place between Fermi arcs at different surfaces\,\cite{Chen2020a}, 
%to the best of our knowledge, no real material has yet been discovered with negative refraction of topological surface states.}
%been neither predicted theoretically nor synthesised experimentally to verify such proposal. 
% In addition to topological semimetals, topological properties have also been extensively studied in other systems such as intrinsic lattice vibrations in solids\,\cite{Stenull2016,Liu2017a,Zhang2018a,Miao2018,Li2018a,Xia2019,Zhang2019c,Liu2020,Li2021,Peng2020a,Wang2020b,Liu2020b,Wang2021,Tang2021,Liu2021,Liu2021a,Liu2021b,You2021,Xie2021,Zheng2021,Wang2021a,Peng2021b,Peng2022,Peng2022a}, but negative refraction of phonon surface states in real materials (rather than artificial phononic crystals) has, to the best of our knowledge, yet to be discovered.
% Material candidates
% Chirality in quartz
% $\alpha$-quartz is the most stable phase of silica under ambient conditions. Other silica polymorphs have structures based on the same tetrahedral unit, sometimes showcasing the same connectivities (like in the case of beta quartz), but often different (as in cristobalite, tridymite, etc...).
Intuitively, negative refraction can take place at the interface between chiral crystals with left- and right-handed screw symmetries, as the isofrequency contours of the surface Weyl arcs are mirror images of each other for a specific choice of orientation. 

One of the most well-known chiral crystal is $\alpha$-quartz, which exists in nature in two enantiomorphs that belong to a space-group pair and are thus handed: the right-handed screw $P3_121$ (No.\,152) and the left-handed screw $P3_221$ (No.\,154). The crystal structures have opposite chirality, which can be distinguished either by measuring their optical activity (the rotation of the plane of polarisation of plane-polarised light), as first observed in quartz crystals in 1811 by Fran{\c{c}}ois Arago\,\cite{Arago1811,Kahr2012}, or by circularly polarised  resonant x-ray diffraction\,\cite{Tanaka2008,Tanaka2010,Igarashi2012}. 
As $\alpha$-quartz is an insulator under ambient conditions, any potential electronic topology away from the Fermi level is not easily accessible. To remedy this, we instead consider the topology of the intrinsic lattice vibrations (phonons) in $\alpha$-quartz, which are not constrained by the Fermi level. The topological properties of phonons have been studied extensively\,\cite{Stenull2016,Liu2017a,Zhang2018a,Miao2018,Li2018a,Xia2019,Zhang2019c,Liu2020,Li2021,Peng2020a,Wang2020b,Liu2020b,Wang2021,Tang2021,Liu2021,Liu2021a,Liu2021b,You2021,Xie2021,Zheng2021,Wang2021a,Peng2021b,Lange2022,Peng2022,Peng2022a}. 
In contrast to metamaterials, phonons are intrinsic quasiparticles in real materials similar to electrons. Additionally, typical phonon frequencies range in 0$-$50 THz, and negative refraction in the terahertz frequency range has been well studied experimentally\,\cite{Hu2023}. Therefore, it is expected that negative refraction can be straightforwardly measured in topological phonons using existing apparatus. The phonon modes in the two enantiomorphs exhibit chiral behaviours such as opposite pseudo-angular momenta, selective optical transitions and opposite transport direction\,\cite{Chen2022a,Zhang2022,Ishito2023a,Tsunetsugu2023,Ishito2023,Oishi2022,Juneja2022,Ueda2023}. Furthermore, band crossings between modes in a single enantiomorph of $\alpha$-quartz form Weyl points because of the lack of inversion symmetry in the space groups $P3_121$ or $P3_221$\,\cite{Wang2020b}, and it is well known that the Weyl points of two enantiomorphs carry opposite Chern number\,\cite{Li2019a,Schroter2019,Sessi2020,Schroter2020,Li2022,Yang2022,Yang2022}. It is therefore expected that the Weyl phonons in $\alpha$-quartz with left- and right-handed screw carry opposite Chern number. 

In nature, quartz naturally forms contact twin structures - two crystals of the same mineral but different orientations, touching along a common plane.
Twinned quartz crystals are much more common than untwinned ones on earth\,\cite{Gault1949}, %. In fact,, many of the two quartz enantiomorphs form contact twins 
%are strongly twinned internally 
and quartz crystals are therefore generally racemic\,\cite{Hazen2003,Lee2022}. In twinned crystals, many different structures with various orientation of the unit cells are possible, and have been generally classified by twinning laws\,\cite{Grimmer2006}. As such, twinned quartz crystals offer a natural and versatile platform to study the interplay between chirality and topology, as the twinning boundaries of $\alpha$-quartz should be easily accessible.
It is worth mentioning that the relationship between chirality and topology is a very active area of research\,\cite{Yan2015,Chang2018,Sanchez2019,Li2020b,Ni2020,Rees2020,Yao2020a,Le2020,Schroter2020,Ni2021,Juneja2021,Bose2021}, as chiral space groups can host a plethora of interesting topological phenomena. Hence, a careful comparison of enantiomorphic structures is of significant current interest, with the potential application of negative refraction occurring at the twin boundary in $\alpha$-quartz.
In this work, we explore the relationship between the twinning type of the chiral crystal structures, their phonon band structure and their associated Weyl points. We show that, for the three most common types of quartz contact twinning, the bulk phonon band structures coincide along all high-symmetry lines in the Brillouin zone. However, depending on the twinning choice, the band structures differ at generic momenta. Furthermore, even when the bulk band structure agrees for enantiomorphic twins, the surface isofrequency contours differ. This allows negative refraction to occur at the twinning interface between two enantiomorphs.

%Interestingly, although the band structures at generic momenta depend on the unit cell choice, the bulk Weyl points remain in the same positions and carry the same Chern number, i.e. the topological properties remain unchanged with Weyl points in different enantiomorphs carrying opposite Chern number.} 

\section{Methodology}

Density functional theory (DFT) calculations are carried out using the Vienna \textit{ab-initio} simulation package ({\sc vasp})\,\cite{Kresse1996,Kresse1996a}. The generalised gradient approximation (GGA) calculations are performed using the Perdew-Burke-Ernzerhof exchange-correlation functional as revised for solids (PBEsol)\,\cite{Perdew2008}, with $4$ valence electrons ($3s^23p^2$) for silicon atoms and $6$ valence electrons ($2s^22p^4$) for oxygen atoms. The plane-wave basis set has an upper kinetic energy limit of $800$\,eV and the $\mathbf{k}$-mesh has a size of $7\times7\times7$, with the self-consistent field loop stopped when energy differences between steps are below 10$^{-6}$ eV.  Structural relaxations are carried out until the Hellman-Feynman forces are less than 10$^{-2}$ eV/\AA.

Density functional perturbation theory is used in calculating Hessian matrices and phonon frequencies\,\cite{DFPT,Gonze1995a}, implemented on a $3\times3\times2$ supercell with a $3\times3\times3$ \textbf{k}-mesh. {\sc phonopy} is used to build the matrix of force constants, diagonalise the dynamical matrix, and obtain the phonon dispersion curves\,\cite{Togo2008,Togo2015}. Convergence of the calculations is assessed by varying both the supercell and $\mathbf{k}$-mesh sizes and noting no discrepant results. The calculations include the splitting between transverse and longitudinal optical phonon modes (LO-TO splitting)\,\cite{Gajdos2006}, but we note that LO-TO splitting plays a minor role on the topological properties of phonons away from the Brillouin centre. {\sc WannierTools}\,\cite{Wu2018} is used to locate every single band crossing point on a phonon $\mathbf{q}$-mesh of size $51\times51\times51$, to calculate the chiralities of the Weyl nodes, and to compute the phonon surface states via the surface Green’s function.

\begin{figure}
\centering
\includegraphics[width=0.35\textwidth]{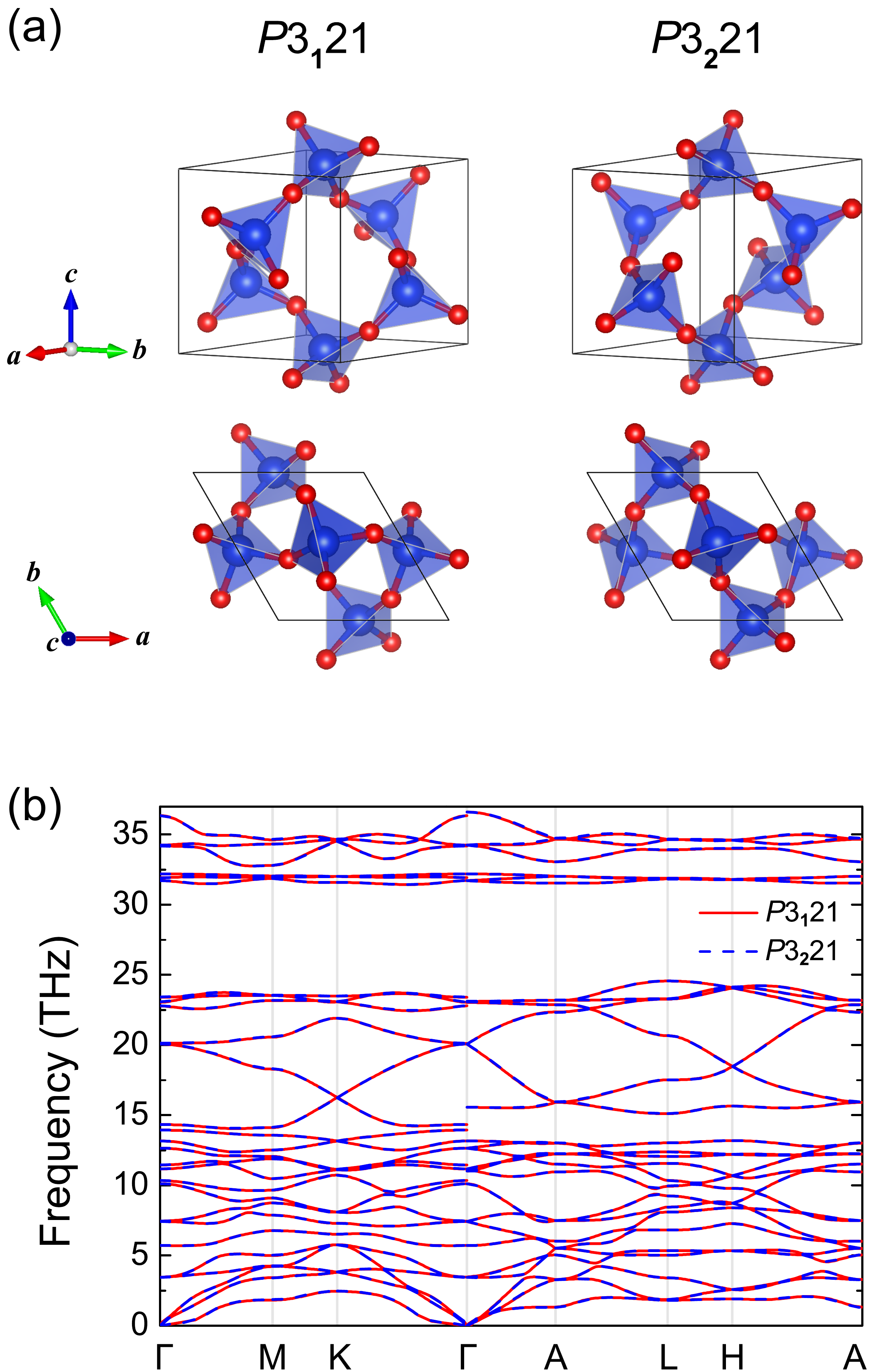}
\caption{
(a) Crystal structures of $\alpha$-quartz with space group $P3_121$ (No.\,152) and $P3_221$ (No.\,154). (b) Phonon dispersion of $\alpha$-quartz. The two structures are related by a mirror operation, leading to Leydolt twinning as explored below.}
\label{crystal} 
\end{figure}

\section{Results and discussion}

\subsection{Crystal structures and lattice dynamics}

$\alpha$-quartz is the most stable phase of silica under ambient conditions, and crystallises in the trigonal crystal system with space group $P3_121$ (No.\,152) or $P3_221$ (No.\,154) depending on the chirality. As shown in Fig.\,\ref{crystal}(a), $\alpha$-quartz is composed of oxygen tetrahedra with silicon atoms placed at their centres. The tetrahedra are joined at their vertices, giving two possible chiral structures. The computed lattice constants are $a=b=4.965$\,\AA\ and $c=5.455$\,\AA, which are in good agreement with previous measured and calculated data\,\cite{Brice1980,Pluth1985,Igarashi2012,Mizokami2018}. We first focus on the conventional unit cell choice, where the ($x$,$y$) position of all atoms in the unit cell agree, and the difference between the enantiomorphs is solely determined by the relative atomic $z$ coordinates, where the enantiomorphs are related by a mirror symmetry, $m_z$, with respect to the $z = 0$ plane. This corresponds to Leydolt twinning, as explored below. Using this unit cell, we show the phonon spectra for $\alpha$-quartz with both space groups in Fig.\,\ref{crystal}(b), agreeing well with previous calculations and measurements\,\cite{Gonze1994,Choudhury2006,Mizokami2018,Wang2020b,Dorner1980}. No imaginary mode is found in the entire Brillouin zone, indicating their dynamic stability. For this choice of twinning, we find that the phonon dispersions of the two chiral structures agrees along all high-symmetry momenta and lines but not at general momenta as shown in Fig.\,\ref{topology}(a) and explored below. The phonon dispersions for other twinning choices are shown in the Supporting Information.

%It is composed of oxygen tetrahedra with silicon atoms placed in their centres, and the tetrahedra join at their vertices, giving two possible chiral structures for alpha quartz. The two enantiomorphs are related by a mirror plane.

% Phonons
\subsection{Weyl points}

% Phonon band crossing points

We focus on the band crossing points formed by phonon branches 18 and 19 in the frequency range 20.5$-$24.0 THz, as they are relatively isolated from the other bands and show the most interesting topological features. On the $q_z = 0$ plane, no band crossing points are formed between these two phonon branches as the two bands are far away from each other. In contrast, the two phonon branches tend to touch on the $q_z = 0.5$ plane (in units of the reciprocal lattice vector 2$\pi$/$c$).

\begin{figure}
\centering
\includegraphics[width=0.4\textwidth]{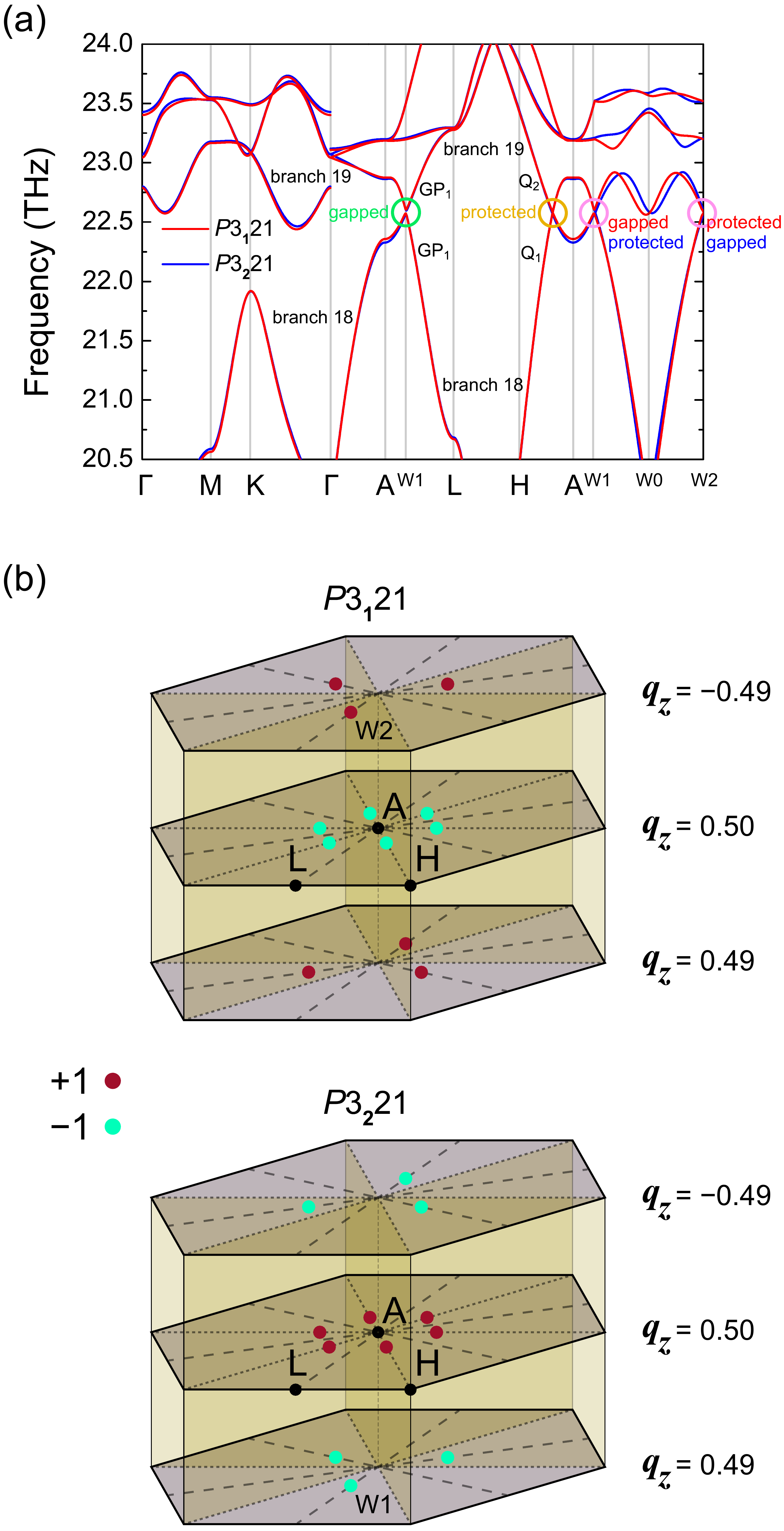}
\caption{
(a) Phonon branches 18 and 19 along the high-symmetry lines and along the line W1 (0.15,\,0,\,0.5)-W0 (0.15,\,0,\,0)-W2 (0.15,\,0,\,$-$0.5) with trivial unitary point-group. (b) Positions of the band crossing points between branches 18 and 19 in the Brillouin zone of $P3_121$ (No.\,152) and $P3_221$ (No.\,154) $\alpha$-quartz, with unit cells related by a mirror operation (Leydolt twinning).}
\label{topology} 
\end{figure}

Along the A-L line, the two bands form an avoided crossing (i.e. they are gapped), as shown in Fig.\,\ref{topology}(a). The point-group along this line is $2'$, so there is no non-trivial unitary point-group symmetry, and hence there is only one irreducible representation (IRREP), GP$_1$, to which both bands belong, as shown in Fig.\,\ref{topology}(a). On the other hand, the point group of the high-symmetry line Q~$=$~H-A is $2$, which gives rise to two different IRREPs Q$_1$ and Q$_2$. We find that bands 18 and 19 belong to different IRREPs along Q and as such, these bands form a stable Weyl point. By threefold rotation and time-reversal symmetry, there are thus a total of six Weyl points on the $q_z = \pm 0.5$ plane, all carrying the same Chern number $\mathcal{C}$ within the same space group. However, for different space groups, the charges $\mathcal{C}$ of the Weyl points on the $q_z = \pm 0.5$ plane are opposite, i.e. $\mathcal{C} = -1$ for $P3_121$ $\alpha$-quartz and $\mathcal{C} = +1$ for $P3_221$ $\alpha$-quartz respectively.

%W1 (0.15,\,0,\,0.5)
%W0 (0.15,\,0,\,0)
%W2 (0.15,\,0,\,$-$0.5)

% Chirality of Weyl points

In addition to the Weyl points on the $q_z = \pm 0.5$ plane, there are also six Weyl points at generic momenta: $\mathcal{C} = +1$ for $P3_121$ $\alpha$-quartz at $(-0.15,0.15,-0.49)$, $(0,-0.15,-0.49)$, $(0.15,0,-0.49)$, $(0,0.15,0.49)$, $(0.15,-0.15,0.49)$ and $(-0.15,0,0.49)$, and $\mathcal{C} = -1$ for $P3_221$ $\alpha$-quartz at $(-0.15,0.15,0.49)$, $(0,-0.15,0.49)$, $(0.15,0,0.49)$, $(0,0.15,-0.49)$, $(0.15,-0.15,-0.49)$ and $(-0.15,0,-0.49)$, respectively [for the chocie of unit cell in Fig.~\ref{crystal}(a) as discussed in the methodology section]. The positions of all the Weyl points in the Brillouin zone are demonstrated in Fig.\,\ref{topology}(b). The six Weyl points at general $\boldsymbol{q}$ in a single enantiomorph are related to each other by the threefold (non-symmorphic) rotation symmetries and time-reversal, neither of which flip the chirality.

\begin{figure}
\centering
\includegraphics[width=0.45\textwidth]{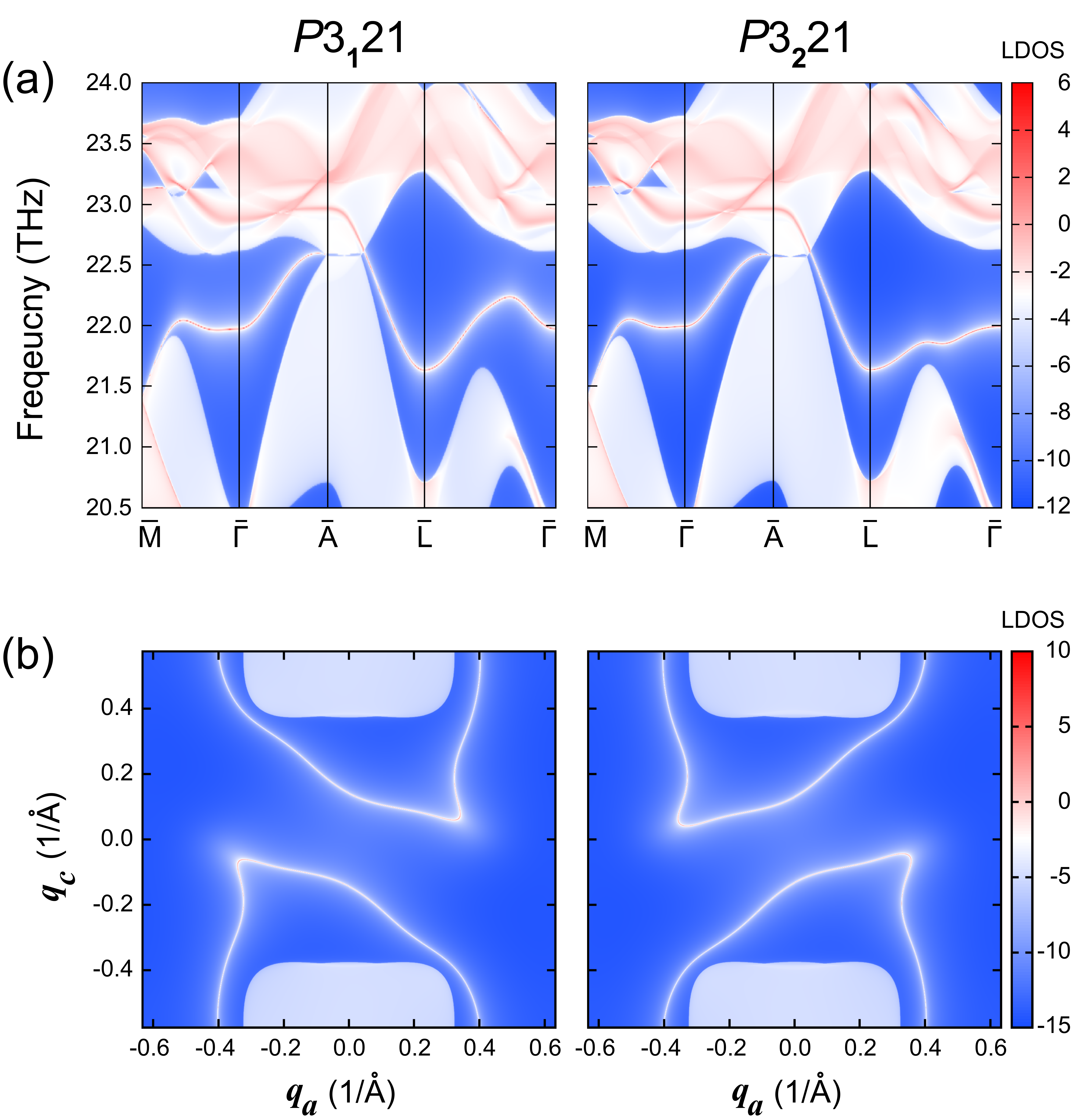}
\caption{
(a) Topological surface states along the [010] direction of $P3_121$ (No.\,152) and $P3_221$ (No.\,154) $\alpha$-quartz with unit cell related by $m_z$ symmetry (Leydolt twinning), and (b) their isofrequency surface arcs at 22.1 THz. Note that the bulk Weyl points are at 22.5 THz so there is no projection of the bulk Weyl points in (b).
}
\label{surface} 
\end{figure}

\subsection{Surface states}
The surface states of the Weyl points projected along the [010] direction are shown in Fig.\,\ref{surface}(a), for the same choice of unit cell as in Fig.~\ref{crystal}(a). The surface local densities of states (LDOS) is calculated from the imaginary part of the surface Green's function\,\cite{Wu2018}. The projections of the bulk Weyl points are connected via surface arcs. The surface arcs along the high-symmetry lines $\bar{\textrm{M}}$-$\bar{\Gamma}$-$\bar{\textrm{A}}$-$\bar{\textrm{L}}$ are exactly the same for both enantiomorphs for this choice of twinning, whereas those at generic momenta along the $\bar{\textrm{L}}$-$\bar{\Gamma}$ are different from each other. For the surface states for other twinning choices, see the Supporting Information.

To have a better view of the distribution of the surface arcs in the reciprocal space, the isofrequency surface arcs at 22.1 THz are plotted in Fig.\,\ref{surface}(b). The surface arcs at a fixed frequency, with the choice of unit cells shown in Fig.\,\ref{crystal}(a) for $P3_121$ and $P3_221$ $\alpha$-quartz, are related by a reflection symmetry along $q_c$. This is analysed further below.

\subsection{Phonon dispersion and twinning}

The space groups $P3_121$ (No.152) and $P3_221$ (No.154) are chiral and form an enantiomorphic pair\,\cite{ITC_A,Nespolo2018}. %Chiral space groups are a subgroup of the Sohncke groups, which are space groups that contain only symmetry operations of the first kind (operations that preserve handedness, i.e. translations, rotations or screw-rotations). Out of the 65 Sohncke groups, 22 (11 pairs) have the special property that their Euclidean normalisers also only contain operations of the first kind, so that the space groups themselves are chiral. 
When the same compound crystallises in two enantiomorphic structures, the unit cells of these structures are related by an operation of the second kind (changing handedness, e.g. mirror, inversion, rotoinversion or glide symmetries).

We will be interested in the interface between these two enantiomorphic structures. We assume that they touch at a crystal plane, e.g. that they from a contact twin along a contact plane. The study of such twins has a long history in mineraology. One distinguishes between merohedral twins (where the lattices are parallel) and non-merohedral twins (where the lattices are not parallel)\cite{ITC_D}. We will only discuss merohedral twinnig here. The three most important merohedral twins of quartz are Dauphiné, Brazil, and Leyoldt (also known as combined-law or Liebisch) twins. Of these, Dauphiné twins occur as twinning between two crystals with the same handedness (same enantiomorphs), whereas Brazil and Leydolt twinning occurs between different enantiomorphs. Each of these is characterised by a different twinning operation. Dauphiné twinning occurs between two crystals whose crystallographic axis are related by a $C_{2z}$ symmetry (twofold rotation around $c$-axis), Brazil twinning occurs between crystals related by $\mathcal{P}$ (inversion through the origin), whereas Leydolt twinning occurs between crystals related by $m_z = \mathcal{P}C_{2z}$ (mirroring in the plane normal to the $c$-axis). In nature, Dauphiné and Brazil twinnings are common, whereas Leydolt twinning is rare\,\cite{ITC_D}. 

The effect of these symmetry operations on the bulk phonon dispersion is given by:
\begin{subequations}
    \begin{align}
    \mathrm{Dauphin\acute{e}:\ }& \omega(q_x,q_y,q_z)&\xrightarrow[]{C_{2z}}\ &\omega(-q_x,-q_y, q_z)\\   
    \mathrm{Brazil:\ }& \omega(q_x,q_y,q_z)&\xrightarrow[]{\mathcal{P}}\ & \omega(-q_x,-q_y, -q_z)\\     
   \mathrm{Leydolt:\ }& \omega(q_x,q_y,q_z) & \xrightarrow[]{m_z}\ &  \omega(q_x,q_y, -q_z).
    \end{align}
\end{subequations}
For the standard unit cell choice in Fig.~\ref{crystal}(a), the crystal structures are related by Leydolt twinning which explains the relative positions of the Weyl points  between the enantiomorphs. As the Berry curvature behaves as a pseudovector, the sign of the Chern number reverses under mirroring. This will also reverse the propagation direction of the topological surface states. We next analyse how the bulk and surface phonon band structures behave under twinning.

\subsubsection{Twinned bulk band structures}
By definition, none of these twinning operations are symmetries of the unitary part of the space groups. However, in non-magnetic systems, time-reversal symmetry $\mathcal{T}$ additionally enforces  $\omega(\boldsymbol{q}) = \omega(-\boldsymbol{q})$. From this, we directly conclude that crystals related by Brazil twinning display the same dispersion for all $\boldsymbol{q}$. 

The same is not true for Dauphiné and Leydolt twinning. We therefore expect the band structure for this twinning to disagree at generic points. To understand the structure at high-symmetry points, we start by noting that both Leydolt twinning and Dauphiné twinning lead to the same constraints on the dispersion, as by time-reversal symmetry, $\omega(-q_x,-q_y,q_z) = \omega(q_x,q_y,-q_z)$. We therefore consider only Leydolt twinning in what follows. We see immediately that on the $q_z = 0$ and $q_z = 0.5$ planes, all twinning types will display the same dispersion. Thus, the only high-symmetry lines left to discuss are $\Delta =$A$'$-$\Gamma$-A and P = H$'$-K-H. Along 
both of these lines, $2_{110}$ symmetry (twofold rotation with respect to the [110] axis) guarantees that the dispersion is symmetric around $\Gamma$ or K respectively, and therefore flipping $q_z\rightarrow -q_z$ does not change the dispersion. Thus, the band structures look identical at all high-symmetry points and along all high-symmetry lines for all twins, as confirmed in Fig.\,S1 in the Supporting Information. This is not true, however, at generic momenta, as shown in Fig.\,\ref{topology}(a). We note in passing that the lower-symmetry chiral space-group pairs $P3_1$ and $P3_2$ do not have $2_{110}$ rotation symmetry, so that in crystals belonging to this chiral space-group pair (twinned by the same operations) we would expect the dispersions to disagree even along the high-symmetry line P ($\Delta$ remains symmetric around $\Gamma$ by time-reversal symmetry).

\subsubsection{Twinned surface states}
The relationship between the surface states for different crystal twins is more subtle, as this depends on both the surface symmetries and on the crystal termination. We focus on the [010] surface with termination for Leydolt twinned crystals in Fig.~\ref{surface}. We find, as shown in Fig.~\ref{surface}(a), that the surface band structures agree along the high-symmetry lines, but differs at generic points. This may be a result of remnant time-reversal symmetry, or of the antiunitary symmetry $2'_{100} = 2_{100}\mathcal{T}$, which leaves $q_y$ invariant. We show the surface band structures for the other twinning operations in Fig.\,S2 in the Supporting Information. More generally, it is not easy to predict the surface symmetries, as these will generally also be influenced by surface termination and reconstruction.

The surface arcs shown in Fig.~\ref{surface}(b) however, arise as a consequence of the bulk Weyl points, and should therefore display some topological protection. We expect these surface states to curve oppositely in the different enantiomorphs, as they are related by a handedness-inverting symmetry, flipping the chirality. This should be the case for both Leydolt and Brazil twinning, as inversion symmetry is generically broken at the surface. Such an inversion of direction in the surface arcs can give rise to negative refraction at the surface of twins.

%This is a direct consequence of the relationship between the materials for this choice of unit cell shown in Eq.\,\eqref{eq:dispersion_enantiomorphs}. We also observe a surface $\bar{q}\rightarrow -\bar{q}$ symmetry separately in each space group, which likely stems from remnant time-reversal symmetry on the surface. 
% Surface states

% Negative refraction
\subsection{Negative refraction}
Benefiting from the isofrequency contours of $P3_121$ and $P3_221$ $\alpha$-quartz, negative refraction can take place at the interface between these enantiomorphs. We focus on the [010] surface for Leydolt-twinned quartz [i.e. crystal structures shown in Fig.\,\ref{crystal}(a)]. 

\begin{figure}
\centering
\includegraphics[width=0.45\textwidth]{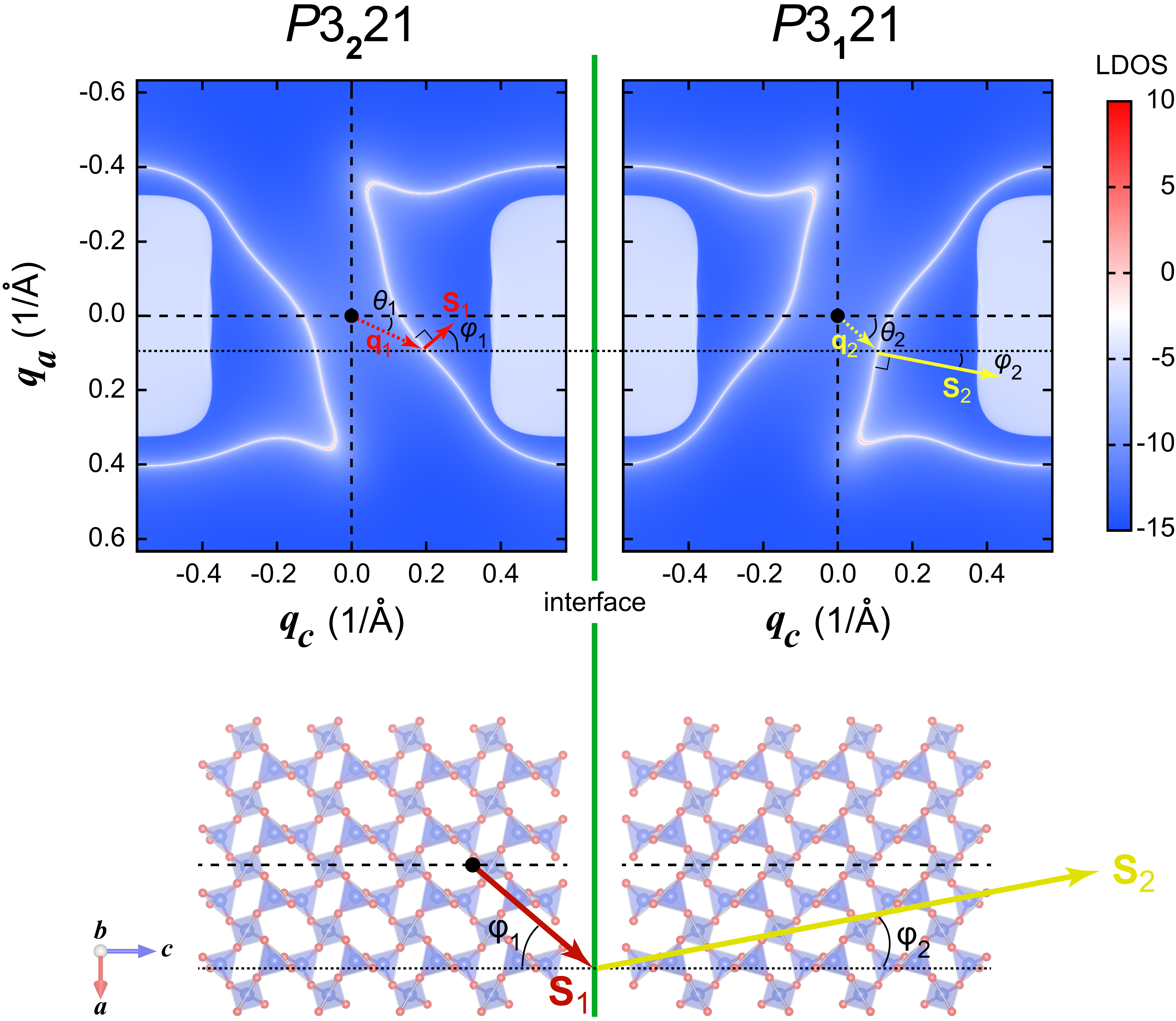}
\caption{
Schematics of negative refraction of surface Weyl arcs along the [010] direction at 22.1 THz, which takes place at the interface between $P3_221$ (No.\,154) and $P3_121$ (No.\,152) $\alpha$-quartz, related by a mirroring operation.
}
\label{negative} 
\end{figure}

We plot the schematics of negative refraction of surface Weyl arcs along the [010] direction that takes place at the interface between the two enantiomorphs in Fig.\,\ref{negative}. For surface phonons at 22.1 THz in $P3_221$ $\alpha$-quartz with wave vector $\textbf{q}_1$ and incidence angle $\theta_1$, the tangential wave vector is conserved\,\cite{Hu2023}, i.e. $\textbf{q}_1 \textrm{sin} \theta_1 = \textbf{q}_2 \textrm{sin} \theta_2$ (where $\textbf{q}_2$ and $\theta_2$ are the wave vector and refraction angle in $P3_121$ $\alpha$-quartz), exhibiting positive refraction for the wave vector. However, the Poynting vector $\textbf{S}$, which is normal to the isofrequency surface arcs and directed along the energy flow, can exhibit negative refraction, as demonstrated by the Poynting vectors $\textbf{S}_1$ and $\textbf{S}_2$ with incidence and refraction angles of $\varphi_1$ and $\varphi_2$ respectively. As a result of negative refraction, the Poynting vectors $\textbf{S}_1$ and $\textbf{S}_2$ of the surface phonons are on the same side of the interface normal. Most interestingly, the negative refraction is tunable by varying the surface phonon frequency (as shown in Fig.\,S3 the Supporting Information). Such negative refraction can transform the linear interface between $P3_221$ and $P3_121$ $\alpha$-quartz into a lens capable of focusing/defocusing phonon waves depending on the frequency-tunable incidence and refraction angles.

In terms of feasibility for synthesising such interfaces, we note that the [010] surface for twinned quartz can be easily found in nature, as quartz twin crystals are much more abundant than the untwinned ones\,\cite{Gault1949}. In virtually every natural quartz, the two morphologically distinct natural crystals are internally twinned, i.e., the enantiomorphs can be found within the same crystal with an interface\,\cite{Hazen2003,Lee2022}. Moreover, the interfaces are experimentally realisable, as it has been reported that an atomically sharp internal interface between two enantiomorphs can be synthesised\,\cite{Mathur2023}.

In terms of measuring negative refraction, previous experiments have used an illumination frequency of about 27 THz by fabricating a gold antenna on one side of the interface to measure the propagation wave\,\cite{Hu2023}, which is similar to the frequency in our work. The refractive behaviour can be measured by a tunable quantum cascade laser in a scattering-type scanning near-field optical microscope\,\cite{Alonso-Gonzalez2014,Li2018c}.

In terms of potential applications, negative refraction in the terahertz region holds significant implications for thermal emission by controlling the flow of thermal energy. This discovery can also facilitate the development of thermal imaging techniques for medical imaging, aerospace and manufacturing.

\section{Conclusion}

We explore the relationship between chiral crystal structures, twinning and topological charges of Weyl points. We find that the, depending on the choice of twinning operation, the bulk band structure for different twins can agree or disagree at generic points. We find further that the Weyl points in opposite chiral structures carry opposite Chern number, which can lead to negative refraction at the twinning surface %for twinning 
between different enantiomorphs.

\section*{Acknowledgements}
G.F.L thanks H. Friis at NORMIN for fruitful discussions. J.D.F.P is grateful for the funding provided by the Peter Mason fund and for the support received from St Catharine's College. G.F.L acknowledges funding from the Aker Scholarship. C.R. acknowledges funding from the Undergraduate Academic Research Projects in St John's College. B.M. and B.P. acknowledge funding from the Winton Programme for the Physics of Sustainability. B.M. also acknowledges support from a UKRI Future Leaders Fellowship (MR/V023926/1) and from the Gianna Angelopoulos Programme for Science, Technology, and Innovation. B.P. also acknowledges support from Magdalene College Cambridge for a Nevile Research Fellowship. The calculations were performed using resources provided by the Cambridge Tier-2 system, operated by the University of Cambridge Research Computing Service (www.hpc.cam.ac.uk) and funded by EPSRC Tier-2 capital grant EP/P020259/1, as well as with computational support from the U.K. Materials and Molecular Modelling Hub, which is partially funded by EPSRC (EP/P020194), for which access is obtained via the UKCP consortium and funded by EPSRC grant ref. EP/P022561/1.

\bibliography{references}
\clearpage
\newpage

\appendix
\onecolumngrid
\section{Supporting Information}

\begin{figure*}[hb!]
	\centering
	\includegraphics[width=0.35\textwidth]{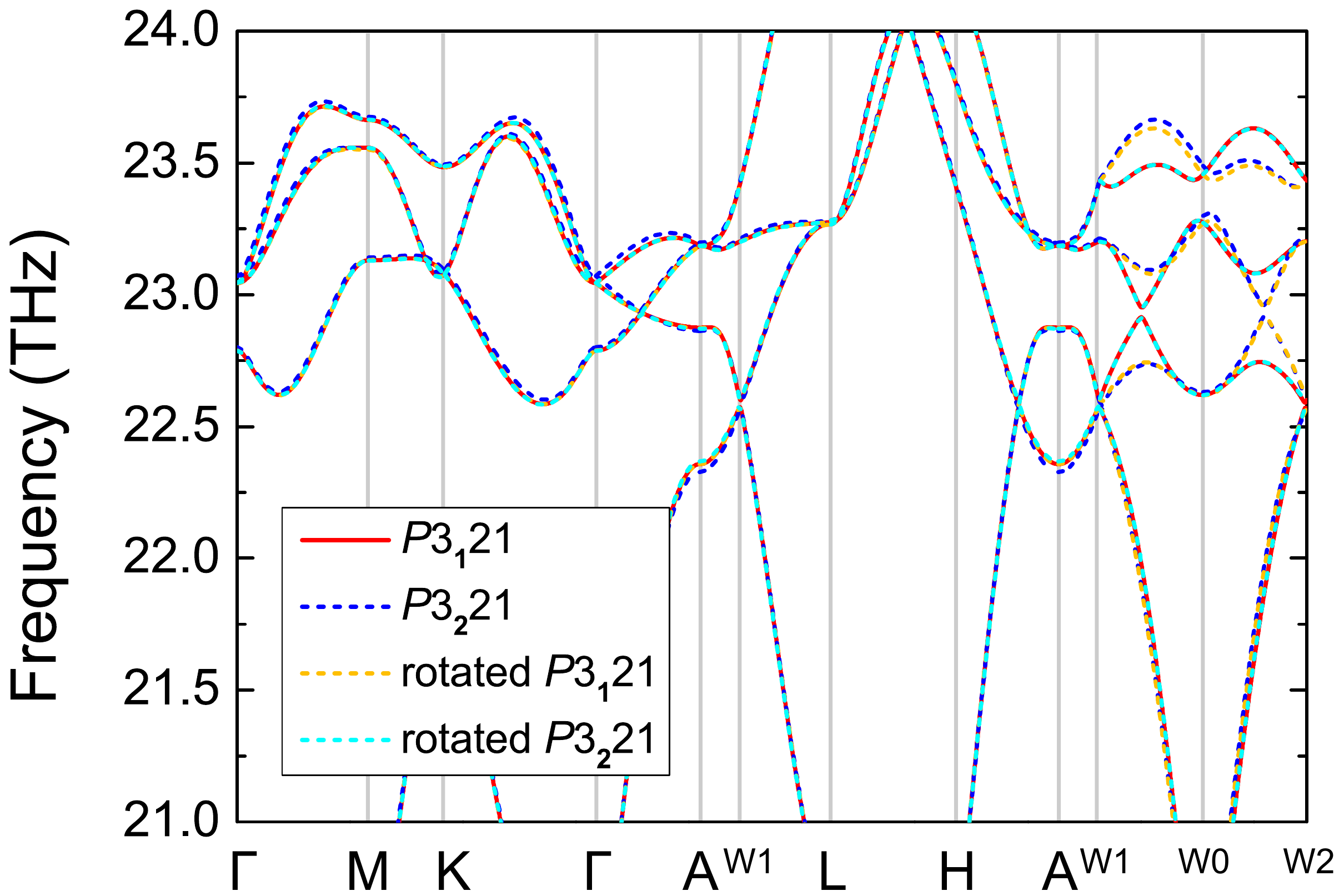}
	\caption{
		Bulk phonon dispersions for four different choices of unit cell, corresponding to different twinning choices. Starting with $P3_121$, $P3_221$ is generated by a mirror-rotation $m_z$ normal to the $c$-axis (corresponding to Leydolt twinning). Rotated $P3_121$ is generated from $P3_121$ by a $2_{001}$ rotation along the $c$-axis (corresponding to Dauphiné twinning relative to $P3_212$). Similarly, rotated P$3_221$ is generated from $P3_221$ by a $2_{001}$ rotation along the $c$-axis (corresponding to Brazil twinning relative to $P3_121$).
	}
	\label{bulk} 
\end{figure*}
\begin{figure*}
	\centering
	\includegraphics[width=0.7\textwidth]{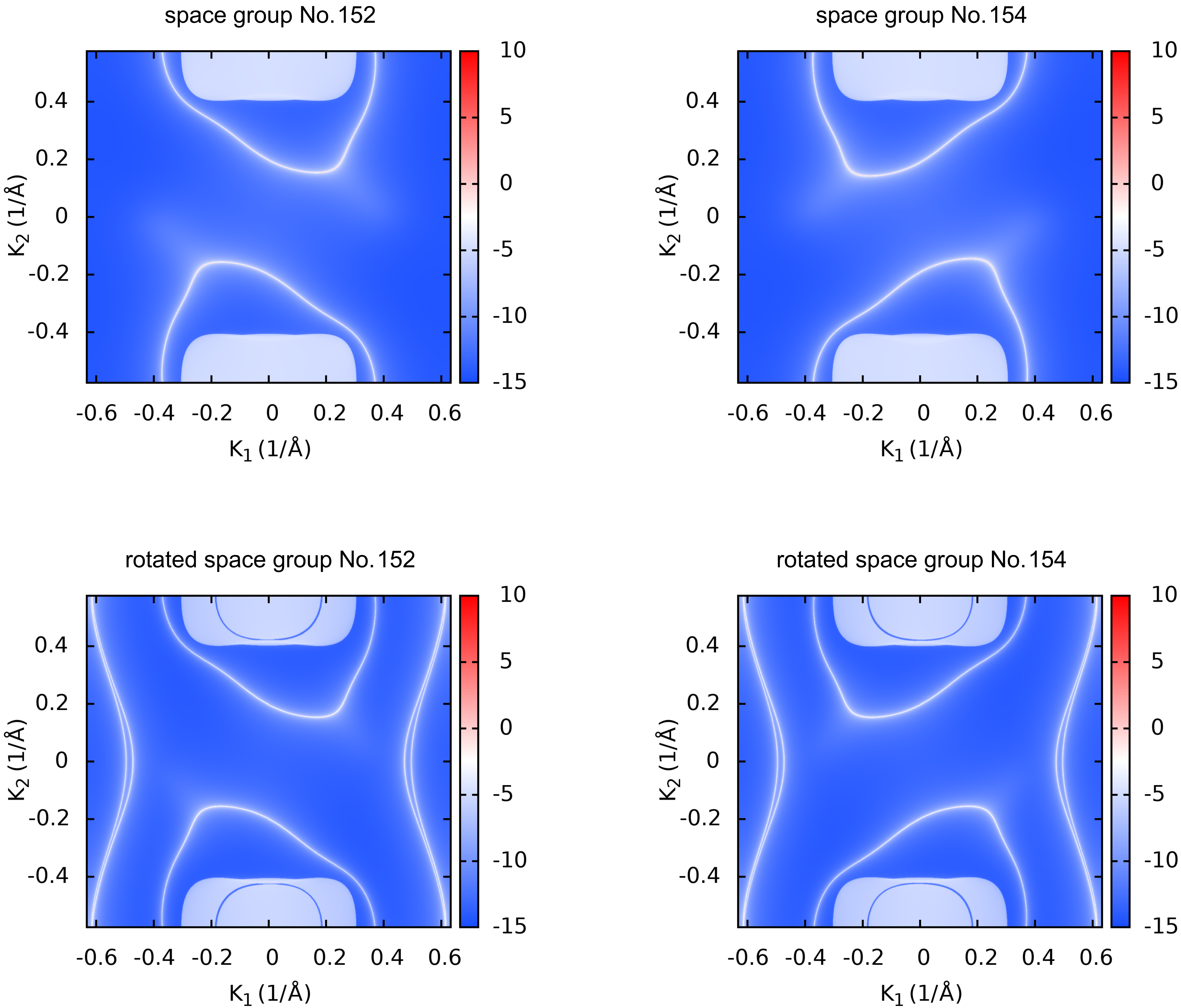}
	\caption{
		Surface arcs for various choices of unit cell for $\alpha$-quartz, corresponding to various twin operations, as described in Fig.~\ref{bulk}. The surface state depends on termination, so that additional surface states can appear upon rotation. Crucially, however, different enantiomorphs curve in opposite direction, as can be seen by comparing the two columns.
	}
	\label{surface} 
\end{figure*}
\begin{figure*}
	\centering
	\includegraphics[width=0.5\textwidth]{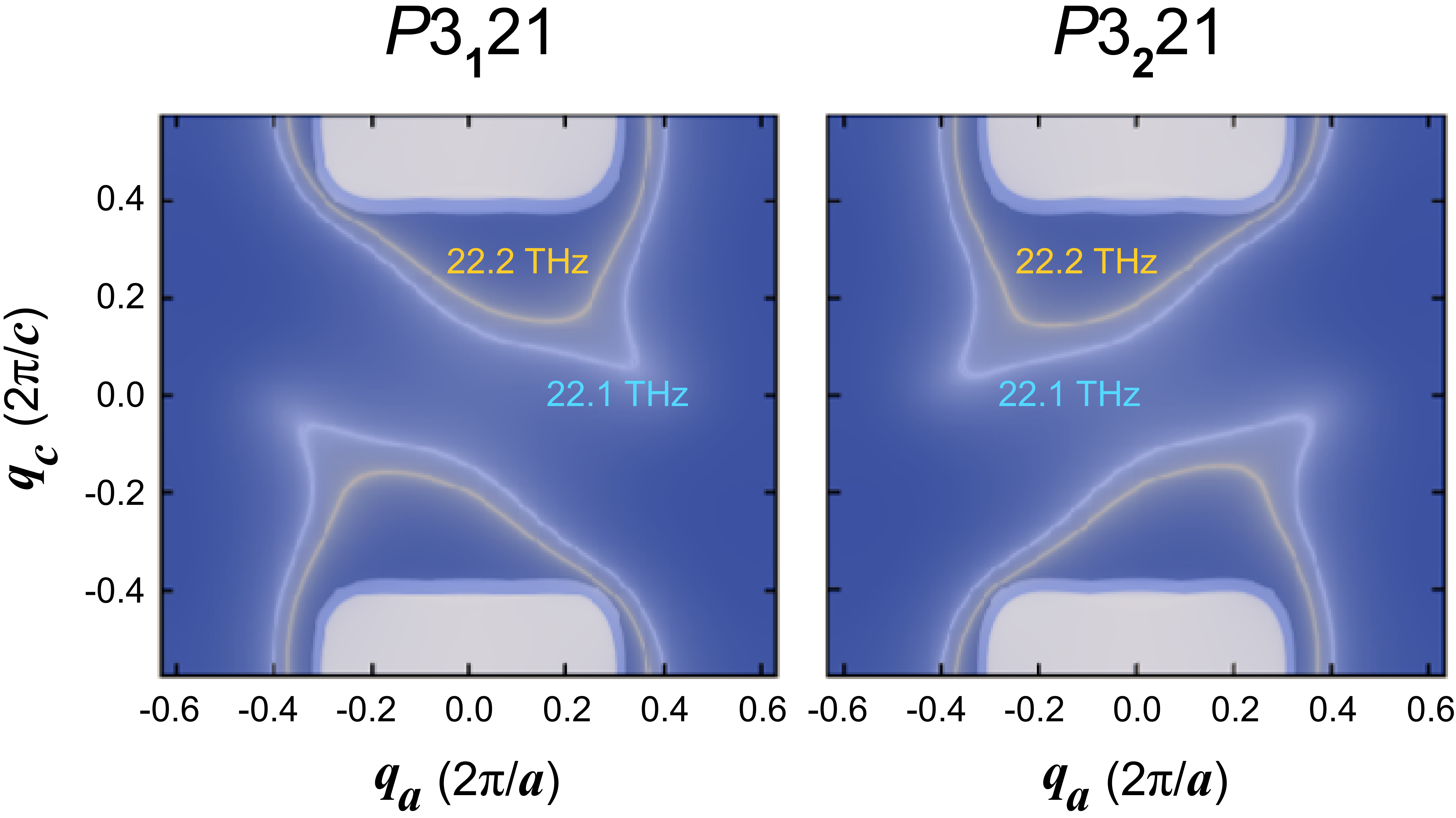}
	\caption{
		Topological surface arcs at 22.1 and 22.2 THz along the [010] direction of $P3_121$ (No.\,152) and $P3_221$ (No.\,154) $\alpha$-quartz related by $m_z$ symmetry (Leydolt twinning).
	}
	\label{tunable} 
\end{figure*}

\end{document}